\documentclass[modern]{aastex63}
\usepackage[utf8]{inputenc}
\usepackage{mathtools}
\usepackage{nameref}
\usepackage{wasysym}
\usepackage{makecell}
\usepackage[caption=false]{subfig}
\usepackage{hyperref}
\usepackage[capitalise]{cleveref}
\usepackage{booktabs}
\usepackage{float}

\graphicspath{{img/}}

\received{July 23, 2020}
\revised{September 14, 2020}
\accepted{September 17, 2020}
\submitjournal{ApJ}

\shortauthors{Simon Müller et al.}

\turnoffedit

\begin{document}

\title{Theoretical vs.~observational uncertainties:\\ composition of giant exoplanets}

\author[0000-0002-8278-8377]{Simon Müller}
\affiliation{Center for Theoretical Astrophysics and Cosmology \\
             Institute for Computational Science, University of Zürich \\
             Winterthurerstrasse 190, 8057 Zürich, Switzerland}
\email{simon.mueller7@uzh.ch}

\author[0000-0002-7355-8318]{Maya Ben-Yami}
\affiliation{Institute of Astronomy, University of Cambridge \\ Madingley Road, Cambridge CB3 0HA, UK}

\author[0000-0001-5555-2652]{Ravit Helled}
\affiliation{Center for Theoretical Astrophysics and Cosmology \\
             Institute for Computational Science, University of Zürich \\
             Winterthurerstrasse 190, 8057 Zürich, Switzerland}
             
\correspondingauthor{Simon Müller}

\begin{abstract}
In order to characterize giant exoplanets and better understand their origin, knowledge of how the planet's composition depends on its mass and stellar environment is required. In this work, we simulate the thermal evolution of gaseous planets and explore how various \edit1{common} model assumptions such as different equations of state, opacities, and heavy-element distributions affect the inferred radius and metallicity. We examine how the theoretical uncertainties translate into uncertainties in the inferred planetary radius and bulk metallicity.

While we confirm the mass-metallicity trend previously reported in the literature, this correlation disappears when removing a 20 $M_{\oplus}$ heavy-element core from all the planets. We also show that using an updated hydrogen-helium equation of state leads to more compact planets. As a result, we present six planets that should be classified as inflated warm Jupiters. We next demonstrate that including the opacity-enhancement due to metal-rich envelopes of irradiated planets changes the planetary radius significantly, which can have large effects on the inferred metallicity. 

\edit1{Even though there are other model assumptions that have not been considered in this work,  we could  show that the calculated theoretical uncertainties can already be comparable or even larger than the observational ones. Therefore, theoretical uncertainties are likely to be even larger}. We therefore conclude that progress in theoretical models of giant planets is essential in order to take full advantage of current and future exoplanetary data.

\end{abstract}

\keywords{planets and satellites: evolution, gaseous planets, interiors, composition --- methods: numerical}

\section{Introduction}\label{sec:introduction}
Investigating how the composition of planets depends on their mass and stellar environment is an important piece of the puzzle in understanding planet formation. The bulk composition of giant exoplanets was first studied by \citet{Guillot2006}, and a possible correlation between the heavy-element content and mass has been suggested thereafter \citep{Miller2011, Thorngren2016}. Since the heavy-element mass of a planet cannot be measured directly, theoretical models have to be used to infer it from mass-radius measurements. In particular, for a given measurement of mass and age, the bulk metallicity needs to be found that reproduces the observed radius.

This linking of theory and observations allowed \citet{Thorngren2016} (hereafter T16) to find a mass-metallicity relation. According to planet formation theories, if most giant planets form by core accretion \citep{Mizuno1980,Pollack1996} they should start with a heavy-element core. As the planet grows, it can gather additional heavy elements via gas accretion, pebbles \citep{Johansen2017} or planetesimals \citet{Mousis2009}. While \citet{Hasegawa2018} showed that the observed mass-metallicity trend can be \edit1{reproduced} \edit1{if the solids are accreted from a planetesimal disk in which the planet has carved a gap.} Recent planet formation models \citep{Venturini2020} have not been able to confirm this scaling. \edit1{In particular, the origin of the apparent high metallicity of some giant exoplanets (e.g., \citet{Thorngren2016,Barragan2018}) is not clear}. One proposed solution is that these planets enhanced their planetesimal accretion by migrating long distances, but even in that scenario it is difficult to explain the extreme metal-enrichment observed \citep{Shibata2020}. An alternative mechanism to explain the mass-metallicity relation is metal-enrichment by major mergers, i.e., collisions of multiple primordial cores \citep{Ginzburg2020}. However, it is currently unclear whether this scenario can explain the many highly enriched gas giants.

Inferring the bulk metallicity of giant planets from mean density measurements is challenging. First, an accurate determination of the age is important, particularly if the planet is young \citep{Fortney2007}. This is because gas giants are mostly composed of hydrogen and helium (hereafter H-He), they are compressible and contract as they cool \citep{Hubbard1977}. Fully convective giant planets contract rapidly within the first $\approx 1$ Gyr, after which the contraction slows down significantly. Second, the planetary metallicity depends on the equation of state (EoS) used by the modeler. Several H-He EoSs currently exist that are used in interior modelling of giant planets (see, e.g., \citet{Saumon1995,Militzer2013,Becker2014,Chabrier2019}). The heavy elements are typically represented by a third material. For the heavy elements, it is common to use either water, some representation of rock (for example, SiO$_2$ or olivine) or a rock-water mixture (see, e.g., \citet{More1988,Thompson1990}). The models also depend on the atmospheric boundary conditions that are imposed, usually in the form of the choice of an opacity and an atmospheric model \citep{Guillot1999,Guillot2010}. Finally, the planet's radius depends on how the materials in the interior are distributed. Due to differences in compressibility of heavy elements and H-He mixtures, previous work has shown that core-envelope structures result in larger radii compared to fully-mixed ones \citep{Baraffe2008,Vazan2013}.

While the influence of these model assumptions on the inferred radius have been investigated independently \citep{Burrows2007,Baraffe2008,Thorngren2016}, we are still lacking a clear picture of the interplay between the various model assumptions and their influence on the model predictions and data interpretation.

Until recently, observational uncertainties of exoplanets were relatively large, and the uncertainties associated with the model assumptions used to infer the bulk composition were mostly negligible. However, as measurements are improving, it becomes essential to understand how the assumption-induced theoretical uncertainties compare to the observational uncertainties, and develop theoretical models that enable a thorough interpretation of the data.

Future exoplanetary atmospheric measurements from NASA's James Webb Space Telescope \citep{Gardner2006} or ESA's Ariel \citep{Tinetti2018} space missions will provide additional constraints. Additionally, the Plato 2.0 mission \citep{Rauer2014} will provide accurate stellar age measurements (5 - 10\%), which will be extremely valuable for the modelling of young giant exoplanets. These stellar ages combined with planetary masses, radii, inferred bulk-metallicities, stellar and atmospheric abundances will help to further constrain various planet formation processes, and provide a statistical overview.

Therefore it is clear that the interpretation of current and future measurements requires strong theoretical foundations and a good understanding of the influence of various model assumptions. 

In this work, we perform a detailed analysis of how different model assumptions affect the inferred radius and metallicity of giant exoplanets. We use a state-of-the-art stellar evolution code to calculate the planetary thermal evolution, and investigate the effect of different EoSs, opacities, and heavy-element distributions. The paper is structured as follows. In \S \ref{sec:methods}, we describe how the thermal evolution is calculated, and present the EoSs and opacities that are used. We proceed with a discussion and independent confirmation of the mass-metallicity relation trend reported in T16 (\S \ref{sec:t16}). In \S \ref{sec:results}, we show that the gas opacity and H-He EoS can have a significant effect on the inferred metallicity of exoplanets. We demonstrate that using an updated H-He EoS \citep{Chabrier2019} yields several warm Jupiters that are inflated in \S \ref{sec:inflated}. Finally, we compare theoretical uncertainties due to model assumptions with observational uncertainties in \S \ref{sec:observations}. Our conclusions are summarized in section \S  6.

\section{Methods}\label{sec:methods}
In order to investigate the effect of model assumptions on the inferred metallicity of exoplanets, we simulate  the thermal evolution of gas giants using the stellar evolution code Modules for Experiments in Stellar Astrophysics (MESA) \citep{Paxton2011,Paxton2013,Paxton2015,Paxton2018,Paxton2019} with a modified EoS suitable for giant planets (see \citet{Mueller2020} for details). We follow a similar approach as \citet{Thorngren2016}: for a given mass, bulk metallicity, and stellar irradiation, MESA calculates how the planetary interior evolves with time. The planet's bulk metallicity is inferred by requiring that its radius at a particular age falls within the observational uncertainty. T16 created a catalogue of non-highly irradiated exoplanets with known mass, radius and age measurements (see \S \ref{sec:t16} for more details), and estimated the bulk metallicity of these planets. Here, we use this catalogue as well as synthetic planets and calculate their thermal evolution with different H-He EoSs, opacities and heavy-element distributions (see \S \ref{sec:mesa} and \ref{sec:opacity} for details).

\subsection{Long-term Thermal Evolution with MESA}\label{sec:mesa}
The modifications from \citet{Mueller2020} to the MESA EoS make it applicable for planetary interiors with heavy-element mass fractions $Z > 0$. This was achieved by calculating the thermodynamic variables for an ideal mixture of hydrogen, helium, and heavy elements, following \citet{Vazan2013}. For H-He, either the SCvH \citep{Saumon1995} or the EoS presented by \citet{Chabrier2019} (henceforth CMS) are implemented. The heavy elements are represented either by water or rock (SiO$_2$) using the QEoS \citep{More1988,Vazan2013}. \edit1{Note that most rocky materials contain iron (e.g., olivine) and have a larger density than SiO$_2$. Therefore, although our EoS is appropriate for the modeling    rocky material in giant planet interiors, our rock EoS  underestimates the density of rock and therefore yield higher metallicity estimates. We briefly discuss this in \S \ref{sec:t16}.}

MESA numerically solves the 1D structure and evolution equations, assuming hydrostatic equilibrium \citep{Paxton2011}, with the Henyey method \citep{Henyey1965}. The 1D grid is spanned by the mass coordinate $m$, and the number of grid points (mass shells) is adaptive. Energy transport by convection, radiative diffusion and electron conduction are included. The dominant energy transport mechanism is determined by applying the Ledoux criterion $\nabla_T < \nabla_{ad} + (\varphi / \delta) \nabla_{\mu}$, where $\nabla_T$, $\nabla_{ad}$ and $\nabla_{\mu}$ are the local temperature, adiabatic and mean molecular weight gradient, respectively. $\varphi = (\partial \ln \rho / \partial \ln \mu)_{P,T}$ and $\delta = (\partial \ln \rho / \partial \ln T)_{P,\mu}$ are material derivatives. Convection is modelled in the diffusive approximation with the mixing-length theory (see, e.g., \citet{Kippenhahn2012}). While MESA is able to self-consistently calculate the mixing of elements, here we disabled material transport by setting the convective diffusion coefficient to zero (\texttt{mix\_factor = 0} in MESA). This allows us to consider stable core-envelope structures \edit1{in addition to fully-mixed ones}, without adding another layer of complication.
\edit1{Giant planets could form with composition gradients in their interior \citep{Helled2017,Valletta2020}, which could delay cooling, in particular if layered convection operates. However, the two extreme cases that we consider in this work are likely to bracket the possible radii at a given age. We discuss and test this assumption in \S \ref{sec:comp_gradients}.}

\subsection{Opacity, Atmospheric Boundary Conditions and Stellar Irradiation}\label{sec:opacity}
The total opacity $\kappa$ is calculated as the harmonic mean of the radiative $\kappa_r$ and conductive opacity $\kappa_{cd}$. For the range of planetary masses that we consider, this effectively means that the relevant opacity for the high pressure-temperature regime in the interior is the conductive opacity, which is calculated from the \citet{Cassisi2007} tables. In the outer envelope where temperatures are lower, conductivity is low and the radiative (gas) opacity becomes relevant. The gas opacity is calculated with the tables from \citet{Freedman2014}, \edit1{which accounts for the opacity enhancement by heavy elements}. The tables provide opacities at metallicities from solar to 50 times solar abundances, assuming the base solar abundances of \citet{Lodders2003} for the equilibrium chemistry calculations. \edit1{Note that these tables do not include a contribution of grains. However, it is still unclear whether grains are generally present in exoplanetary atmospheres over relevant timescales, or whether they quickly settle (e.g., \citet{Movshovitz2010,Mordasini2014}). In \S \ref{sec:observations}, we discuss the implications on the theoretical uncertainty of this simplification.}

The planet's radius $R$ is determined by the location of the photosphere, i.e.,  where the optical depth $\tau = \int_{R}^{\infty} \kappa \rho dr = 2/3$, where $\rho = \rho(r)$ is the density at a given radial coordinate $r$. The location of the photosphere strongly depends on the atmospheric opacity, and as such the inferred planetary radius at a given mass and age is affected by the gas opacity. 

While strongly irradiated and inflated planets are excluded from the T16 catalogue, the planets still receive significant stellar flux of $F_{*}\sim 10^{5} - 10^{8}$ erg cm$^{-2}$s$^{-1}$. In order to account for the heating of the atmosphere due to stellar irradiation, we used the incident stellar irradiation to define an energy generation rate $\epsilon = F_{*} / 4 \Sigma_{*}$. The energy from the irradiation was then applied to the outer mass column $\Sigma \leq \Sigma_{*}$ \citep{Paxton2013}, where $\Sigma(r) = \int_{r}^{R} \rho(r^{\prime}) dr^{\prime}$. This represents the absorption of stellar optical radiation below the photosphere. We use the column depth $\Sigma_{*} = 3 \times 10^{2}$ g cm$^{-2}$, and set the corresponding values for the \texttt{irradiation\_flux} and \texttt{column\_depth\_for\_irradiation} MESA inlist options.

We use the \texttt{which\_atm\_option = "simple\_photosphere"} setting, which means that the location of the photosphere is
estimated for $\tau = 2/3$ without integration. This model atmosphere works well for stellar fluxes $F_{*} \leq 10^{9}$ erg cm$^{-2}$ s$^{-1}$, and is appropriate for the warm Jupiters we consider in this work. The \texttt{simple\_photosphere} option yields good convergence over a large range of planetary masses, compositions and stellar fluxes. For the modelling of strongly irradiated planets (hot Jupiters), MESA provides the atmospheric model from \citet{Guillot2011} (\texttt{which\_atm\_option = "gray\_irradiated"}). It should be noted that at very high fluxes the Freedman opacity tables used by MESA are no longer valid. This is because these opacities are calculated using the local gas temperature in the weighing function of the Rosseland mean. For strongly irradiated atmospheres, the stellar effective temperature should be used, which can lead to dramatically different values for the opacity (see \S 5 in \citet{Freedman2014} for details).

\section{Discussion of the T16 Data, Assumptions and Results}\label{sec:t16}
The T16 catalogue includes warm giant planets with known masses and radii, typically determined by radial velocity and transit measurements, respectively. Since the inflation mechanism of Hot Jupiters is still unknown (see, e.g., \citet{Fortney2010,Weiss2013,Baraffe2014} for a review), planets that receive a stellar flux beyond $2 \times 10^{8}$ erg cm$^{-2}$ s$^{-1}$ are excluded. Thus, the mass-metallicity relation of planets should not be affected by the unknown inflation mechanism. The masses of the planets in their catalogue are within the range of $\simeq 0.01 - 10 M_{J}$.

For the thermal evolution, T16 use their own planetary evolution code that solves the usual 1D structure equations in hydrostatic equilibrium. For the EoS, they use the ideal mixing law to combine the \citet{Saumon1995} EoS for H-He with the ANEOS EoS \citep{Thompson1990} for a 50\%-50\% rock-ice mixture. Note that rock in ANEOS is represented by the mineral olivine and not by SiO$_2$ as in QEoS. Since SiO$_2$ is less dense than olivine, a 50\%-50\% rock-ice mixture calculated with ANEOS yields roughly the same density as pure SiO$_2$ in our EoS. For their atmosphere models, T16 interpolated on the solar metallicity grids from \citet{Fortney2007}. As discussed in T16, this is not fully self consistent, since most of these planets are expected to have super-solar atmospheric metallicities, which would slow their cooling \citep{Burrows2007} and result in larger estimated heavy element masses. The heavy elements were distributed by placing up to 10 $M_{\oplus}$ into a pure heavy element core, while the rest were homogeneously mixed into the envelope. With their assumptions for the atmosphere, a heavy-element core and a H-He envelope with no heavy elements yields larger radii compared to a homogeneously mixed planet. This is because H-He mixtures are more compressible than typical heavy elements \citep{Baraffe2008}.

\begin{figure}[h]
    \epsscale{1}
    \plotone{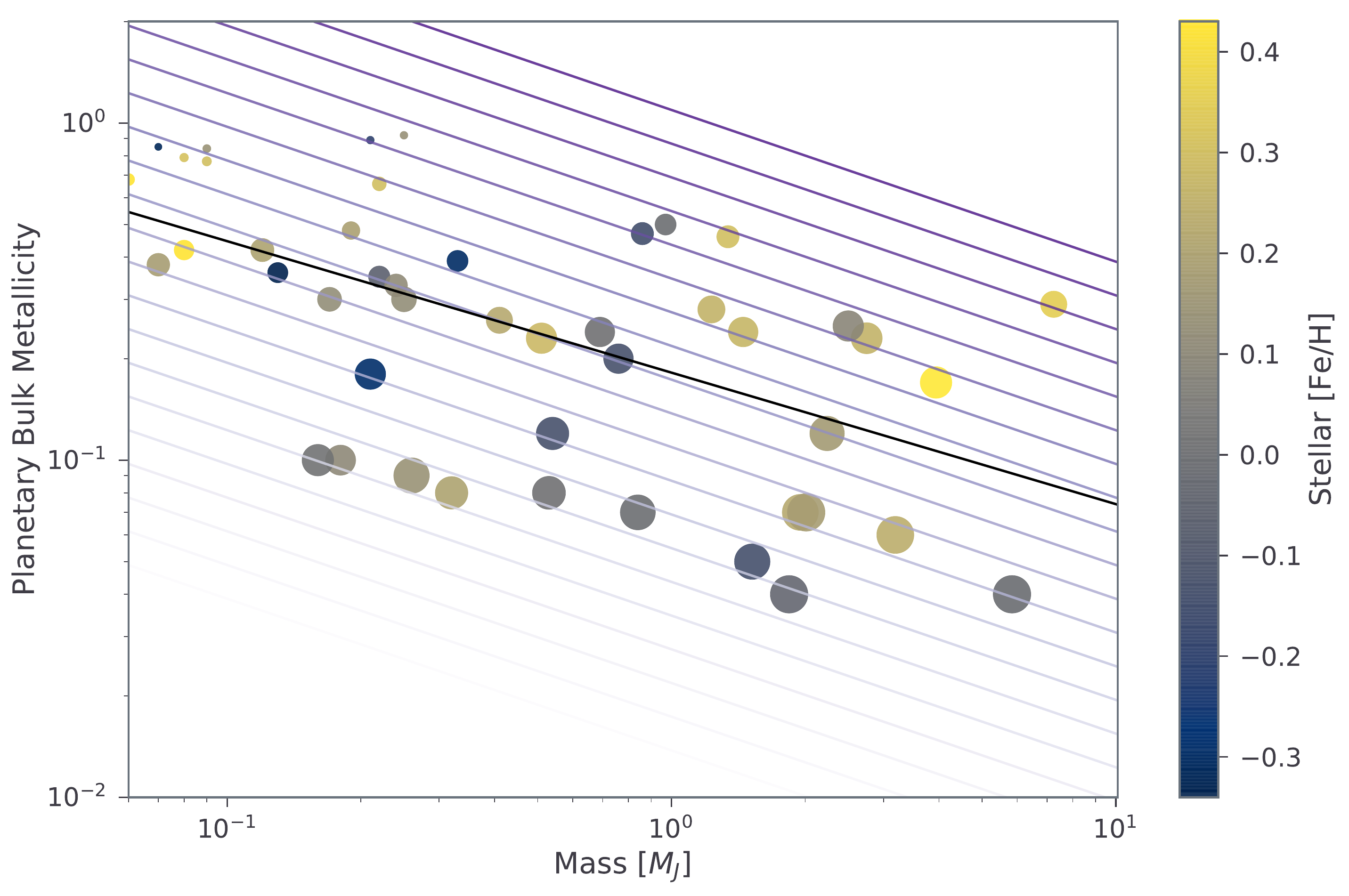}
    \caption{Planetary bulk metallicities (as inferred in T16) plotted against the observed planetary masses. The black line shows the mass-metallicity fit from T16. The purple lines in the background show the $Z \propto Z_* M^{-0.45}$ fit for different values of stellar [Fe/H] in the range of -1 to 1. The scatter points are coloured by the metallicity of the host star, and their size is scaled with the planetary radius.}
    \label{fig:t16_zp_fit}
\end{figure}

Since our code and model assumptions (e.g., EoS, atmosphere, opacity) are different than those used by T16, we first investigated whether we could roughly reproduce their mass-metallicity trend. We used the T16 results as a starting point and ran models for each planet using the median mass and the predicted planetary metallicity $Z$ together with its lower- and upper- bounds. We used SCvH for H-He and SiO$_2$ as the stand-in for the heavy-element, as discussed above. Furthermore, we forced MESA to use solar metallicity in the opacity calculation to more closely mirror the T16 models. While our results were slightly different, we recovered a very similar mass-metallicity relation. 

% $M_z = (57.9 \pm 7.03) M^{0.61 \pm 0.08}$
% $Z / Z_* = (9.7 \pm 1.28) M^{-0.45 \pm 0.09}$

The mass-metallicity relation given in T16 is approximately $M_z \propto M^{0.6}$, where $M_z$ and $M$ are the heavy-element and the total planetary mass, respectively. Using the definition that $Z =  M_z / M$, they also give a relation for the ratio of planetary to stellar metallicity (the heavy-element enrichment), which is roughly $Z / Z_* \propto M^{-0.45}$.

In \cref{fig:t16_zp_fit} we show the planetary mass-metallicity scatter (coloured points) together with the T16 fit (black line). The purple shaded lines in the background show the $Z \propto Z_* M^{-0.45}$ fit for a range of stellar [Fe/H], where the approximation $Z_* = 0.014 \times 10^{[Fe/H]}$ was used. T16 discussed that there may be a connection between the planetary and stellar metallicity (see their \S 5.3 and \S 6) but it does not follow a simple power-law. This is also demonstrated in \cref{fig:t16_zp_fit}: if there were a power-law relation between the planetary and stellar metallicity, planets whose stars have the same [Fe/H] (dots with the same colour) should lie on parallel lines in the $(M, Z)$ plane.

The plot clearly shows that planets on a line predicting similar associated stellar [Fe/H] show no correlation with the stellar metallicity. For example, on the line for [Fe/H] $= 0.2$ the actual values for the planets range from $-0.3$ to $0.4$. This result is consistent with the recent analysis of planetary metal enrichment by \citet{Teske2019}, where no clear correlation between the host star and planetary residual metallicity \edit1{(the relative amount of metal versus that expected from the planet's mass according to the T16 fit)} was found.
\edit1{This result might not be surprising, since the heavy-element mass in planets is not expected to come from the accretion of stellar-composition gas. It is interesting to note that two of the T16 planets orbit the same star (Kepler 30). Their inferred residual metallicites are 0.48 for Kepler-30c and 0.75 for Kepler-30d, respectively. Therefore, their heavy-element masses, beyond what is expected from the T16 mass-metallicity relation, are sub-stellar and are not correlated with the host star. This trend is also valid  for the planets in our solar system: the residual metallicities of Jupiter and Saturn are rather different, with the exact values depending on the heavy-element estimates.}

% Kepler-30c: M = 2.01 Mj, Mz = 42.54 Me, Residual Metallicity = 0.479
% Kepler-30d: M = 0.07, Mz = 8.70, Residual Metallicity = 0.75

\begin{figure}
    \epsscale{1.15}
    \plotone{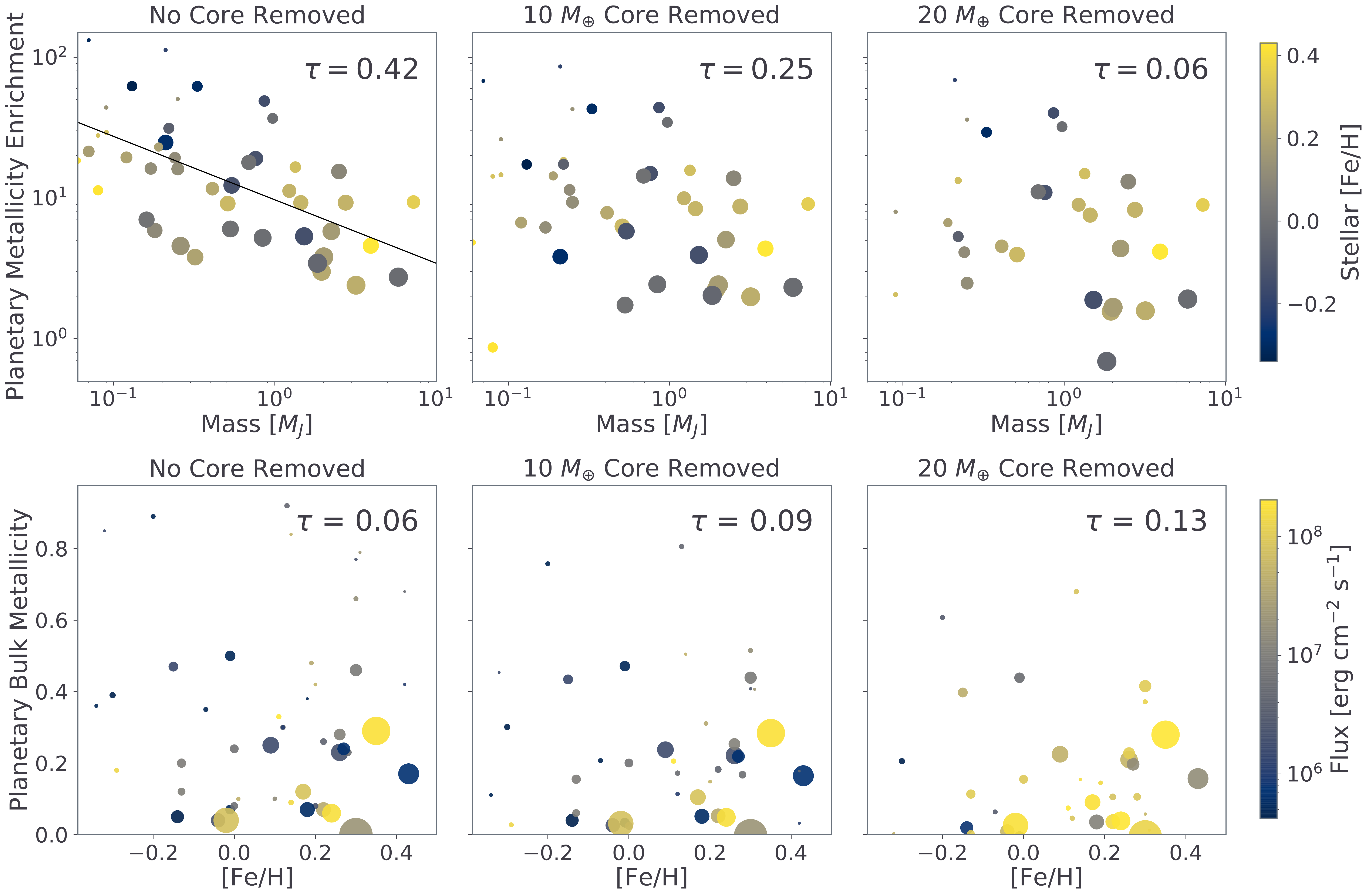}
    \caption{\textbf{Top:} Metal enrichment ($Z / Z_*)$ of the planets in the T16. From left to right: the original data, effective metal enrichment by removing a $10 M_{\oplus}$ and $20 M_{\oplus}$ heavy-element core. The black line in the first figure shows the T16 fit. The scatter points are coloured by the metallicity of the host star, and their size is scaled with their radius. \textbf{Bottom:} Planetary metallicity ($Z$) vs. stellar [Fe/H] of their host stars from the T16 catalogue (left), and with a $10 M_{\oplus}$ (middle) and $20 M_{\oplus}$ heavy-element core removed (right). The sizes of the scatter points correspond to their mass, and the colour scale indicates the stellar flux they receive. In each panel Kendall's tau rank correlation coefficient is shown.}
    \label{fig:t16_nocore}
\end{figure}

\citet{Hasegawa2018} analysed and interpreted the T16 results, and examined the effect of solid accretion on the planetary metallicity by subtracting possible heavy-element core masses from the the total heavy-element content (see their \S 4.3). Here, we followed a similar approach: we took the exoplanets of the T16 catalogue, together with the best estimate for their metallicity, and subtracted values for the core mass that were larger (10 and 20 $M_{\oplus}$) than in \citet{Hasegawa2018} (1, 5 and 10 $M_{\oplus}$). The larger 10 - 20 $M_{\oplus}$ core mass could, for example, correspond to the pebble isolation mass (see, e.g., \citet{Bitsch2018}).

We present the resulting $Z / Z_*$ vs.~$M$ relation in the upper panels of \cref{fig:t16_nocore}. We calculated Kendall's tau rank correlation coefficient for all the cases (no core mass removed, $10 M_{\oplus}$ and $20 M_{\oplus}$ removed), yielding $\tau = 0.42, 0.25, 0.06$. Note that 14 planets have less than $20 M_{\oplus}$ of heavy elements, and we removed these from the correlation calculation and the figure. As \citet{Hasegawa2018} noted, the removal of the core mass yields a much shallower slope than $Z / Z_* \propto M^{-0.45}$. Interestingly, once a $20 M_{\oplus}$ core is removed, there appears to be no correlation at all between $Z / Z_*$ and the planetary mass $M$ (p-value: 0.69). The same effect occurs for the correlation between $Z$ and $M$. Therefore, if most giant planets have a $20 M_{\oplus}$ (or larger) core, there is no correlation between planet mass and heavy-element content.

The core masses predicted by standard formation models have a large range, between $\simeq 1 - 20 M_{\oplus}$. The actual value depends on the core accretion rate, the disk model and the physics included in the calculation \citep{Helled2014}. For pebble accretion, the pebble isolation mass represents the maximum core mass, which is around $\simeq 20 M_{\oplus}$ \citep{Lambrechts2014,Johansen2017}. Recent studies indicate that most accreted solids do not reach the central core, but are expected to ablate \citep{Valletta2019} or evaporate \citep{Alibert2017,Brouwers2018}. It is an ongoing investigation whether the heavy elements sink to the core, remain in a stable composition gradient, or are mixed into the envelope by large scale convection (see, e.g., \citep{Vazan2015,Vazan2016,Vazan2018,Mueller2020}). While there is likely no standard core mass for giant planets, massive cores are plausible.

In the bottom panels of \cref{fig:t16_nocore} we show the planetary bulk metallicity vs.~ stellar [Fe/H].
The vanishing correlation of $Z$ and $M$ once a $20 M_{\oplus}$ core is removed implies that the mass-metallicity relation arises only because envelope-to-core mass ratio changes with the planetary mass. Assuming that the gaseous envelope has a similar metallicity \edit1{to} the proto-planetary disk and the star, the envelope metallicity is expected to be correlated with \edit1{the} stellar [Fe/H]. Kendall's tau increased by a factor of two when removing a $20 M_{\oplus}$ core, indicating a weak correlation between the two variables. However, the result is not statistically significant and there is about a 1/3 chance that the correlation occurs by chance (p-value: 0.29).

\edit1{The gaseous envelope does not need to have the same composition as the host-star. It could, for example, be enriched by upward mixing of heavy elements from a primordial composition gradient \citep{Vazan2015} or core erosion  \citep{Guillot2004}. In addition, the envelope could be polluted from the formation process itself. Depending on the size and composition of the accreted planetesimals/pebbles, they are unlikely to reach the core and are evaporated in the envelope (e.g., \citet{Valletta2019, Valletta2020}).  Late-stage planetesimal during runaway gas accretion may also further pollute the envelope with heavy elements \citep{Podolak2020}. These effects could lead to super-stellar envelope metallicites.}

\section{Influence of Model Parameters on the Inferred Metallicity of Exoplanets}\label{sec:results}
\subsection{The H-He EoS}\label{sec:eos}
Here, we investigate the influence of two different H-He EoSs on the planetary inferred radius. For the comparison we used the widely used SCvH \citep{Saumon1995} EoS, as well as its updated version CMS \citep{Chabrier2019} (see \S \ref{sec:methods}). For all the cases, we keep the H-He ratio at the proto-solar value ($X_{proto} / Y_{proto} = 2.74$).

\begin{figure}
    \epsscale{1}
    \plotone{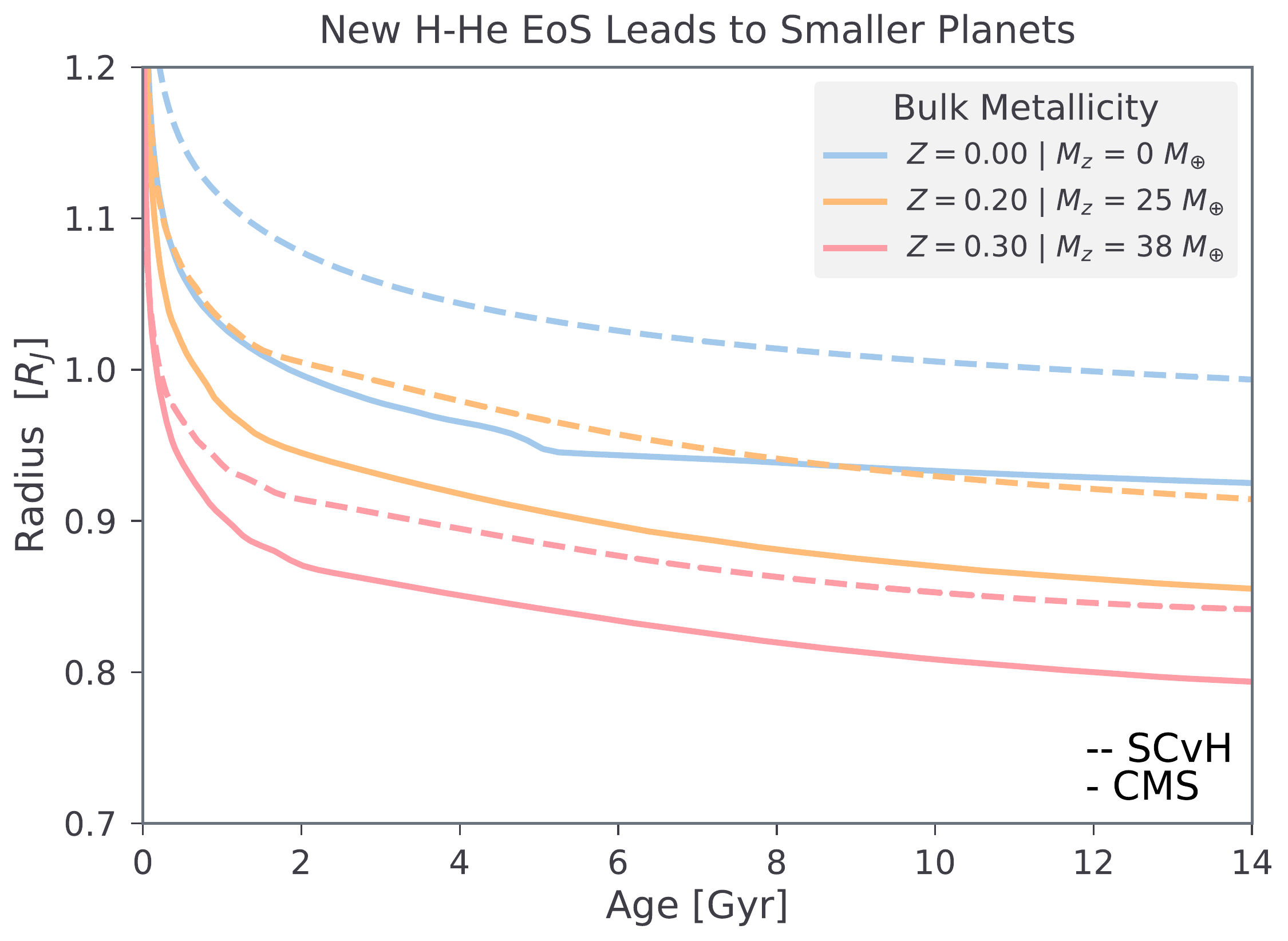}
    \caption{Planetary radius vs.~ time for a 0.4 $M_J$ planet calculated for three different mass fractions of water homogeneously mixed in the interior. The dashed lines use the SCvH EoS for H-He, while the solid lines use CMS. The colours indicate the bulk metallicity and the respective heavy-element masses are as noted in the legend.}
    \label{fig:cms_scvh_mixed}
\end{figure}

\cref{fig:cms_scvh_mixed} we shows the radius evolution of a 0.4 Jupiter mass ($M_J$) planet for three different mass fractions of water (0, 0.2 and 0.3) and a homogeneously mixed interior. The synthetic planet receives a stellar flux of $10^{7}$ erg cm$^{-2}$ s$^{-1}$, and we allow the gas opacity to scale with the envelope's metallicity.

At a given age and metallicity, the CMS EoS yields a planet that is more compact than the one calculated with SCvH. This is because hydrogen is denser under certain pressures and temperatures in CMS in comparison to SCvH \citep{Chabrier2019}. As a result, the effect of using these two EoSs is more profound for lower $Z$ values, where the hydrogen mass fraction is larger. Note that the zero metallicity line for CMS is very close to overlapping with the $Z = 0.2$ line for SCvH. This shows that the change in radius caused by only changing the H-He EoS can induce a large change in the determination of the planet's metallicity.

\begin{figure}[h]
    \plotone{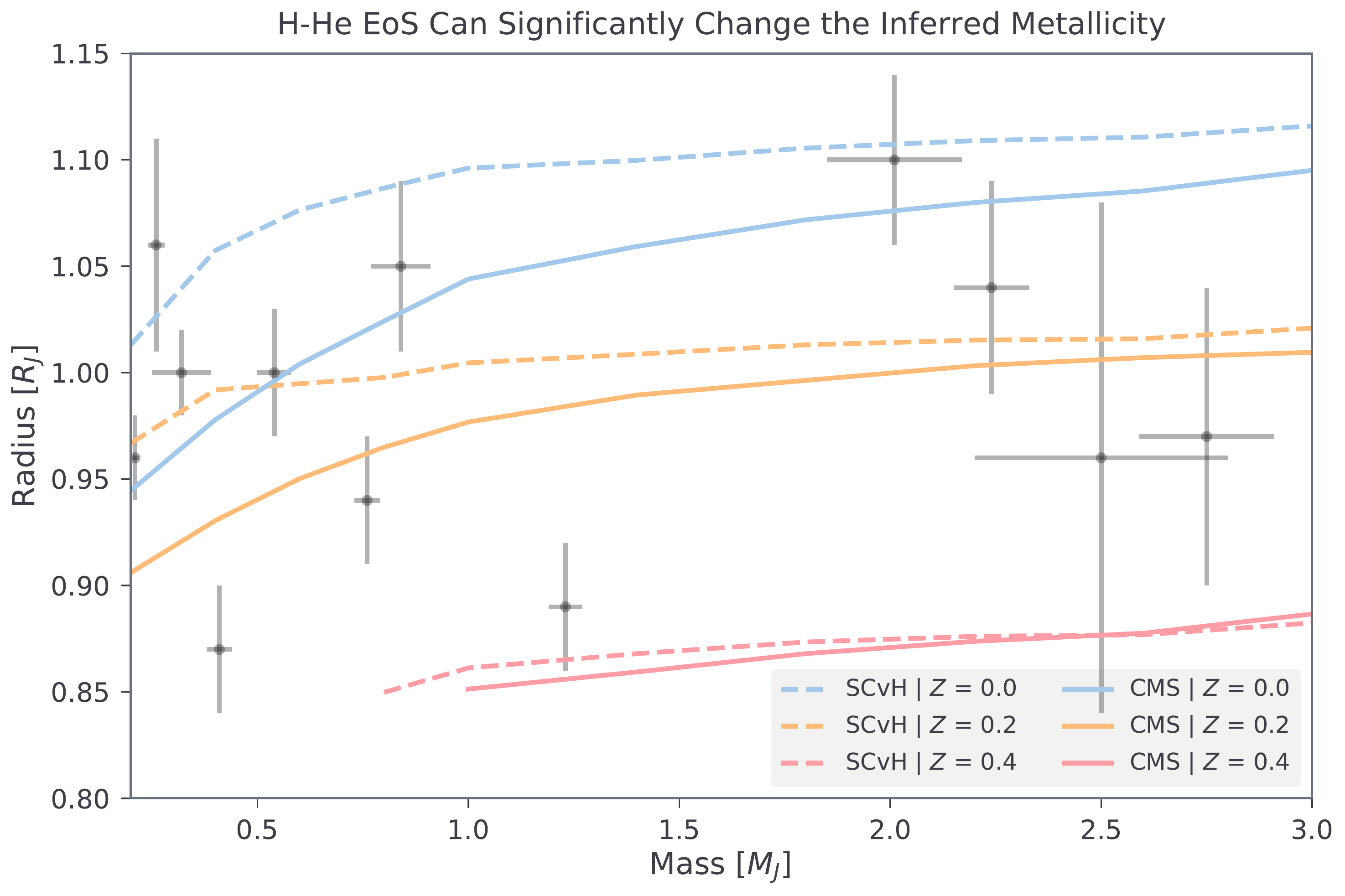}
    \caption{Planetary radii as a function of mass at 3 Gyrs. Shown are planets with masses between $0.2 - 3 M_J$. Solid and dashed lines are calculated with CMS and SCvH, respectively, and the colours indicate the planetary bulk metallicity. For comparison, the gray error bars show observed exoplanets and their measurement uncertainties.}
    \label{fig:mass_radius_3gyrs}
\end{figure}

Next, for each of the EoSs, we calculated the thermal evolution of synthetic planets until the age of 3 Gyrs on a mass-metallicity grid. The masses range from $0.2 - 3 M_J$ and the planetary bulk metallicity is either $Z = 0, 0.2$ or $0.4$. The heavy-element distribution, the flux and opacity are the same as in the previous case. In \cref{fig:mass_radius_3gyrs} we plot the calculated radius at 3 Gyrs vs.~planetary mass for SCvH (dashed lines) and CMS (solid lines). Again, it is clear that the difference in radius is more profound when the bulk metallicity is low.

In addition, the effect is more significant with decreasing mass. At masses $< 1 M_J$, the difference can be as large as 10\%, while beyond $\simeq 2 M_J$ it is at most a few percent. The $Z = 0.2$ lines calculated with CMS and $Z = 0$ calculated with SCvH intersect at $\simeq 0.5 M_J$. The change in radius is sufficient to significantly influence the inferred planetary metallicity.

The key conclusion from this figure is that even if the mass, radius, and age of a giant planet are accurately determined, the inferred metallicity could vary by up to 20\% when different H-He EoSs are used.
\clearpage

\subsection{Opacity}\label{sec:opacity_results}
In this section we investigate how the change in envelope metallicity and therefore gas opacity affects the inferred radius for a given mass, planetary bulk metallicity, and stellar irradiation.

As discussed in \S \ref{sec:opacity} the \citet{Freedman2014} opacity accounts for the heavy elements present in the gas. Over the temperature range from 250 K to 3000 K, which is relevant for many observed exoplanets, the gas opacity scales strongly with metallicity, increasing almost linearly. At lower temperatures, the \edit1{gas} opacity only scales weakly with metallicity (see Fig. 6 in \citet{Freedman2014}). \edit1{Note that this would not be the case if the contribution of grains were included in the opacity. This could significantly contribute to the total opacity and delay the cooling for low atmospheric temperatures.}

Since most of the observed exoplanets have ages of a few $10^{9}$ years and are thought to have cooled down, their atmospheric temperature is in large part determined by the incident stellar irradiation \citep{Marley2007}. As a result, there is an interplay between the atmospheric metallicity, the opacity, and the stellar irradiation. On the other hand, in theoretical models the atmospheric metallicity is determined by the assumption of how the heavy elements are distributed in the interior. For example, if all the heavy elements are in the core, then the atmospheric metallicity and consequently the opacity is lower. This lower opacity causes a faster heat loss and therefore results in a smalle rplanet at a given age.

We calculated the thermal evolution for two planets with $M = 0.2 M_J$ and $1 M_J$, respectively. For each planet, we simulated the evolution assuming four different stellar irradiation fluxes: no irradiation, $10^{5}$, $10^{7}$ and $10^{8}$ erg cm$^{-2}$s$^{-1}$. The heavy elements were either all in the core (no metals in the atmosphere), or homogeneously mixed throughout the planet (metals in the atmosphere).

\begin{figure}
    \epsscale{1.15}
    \plotone{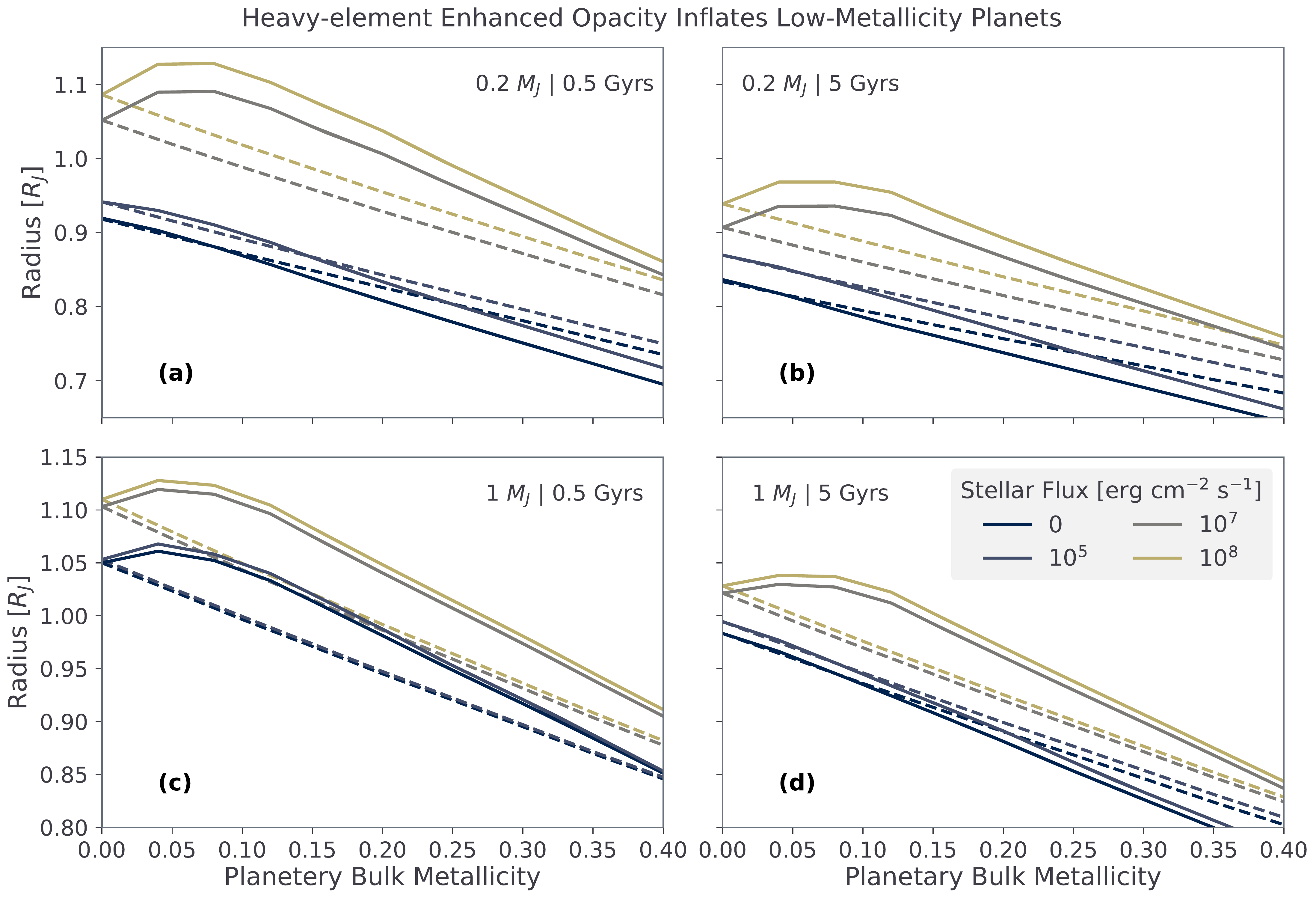}
    \caption{Radius vs. planetary bulk metallicity of a $0.2 M_J$ (top) and a $1 M_J$ (bottom) planet at 0.5 (left) and 5 Gyrs (right). All the heavy elements are placed in a core (dashed lines) or homogeneously mixed (solid lines). The colors indicate the stellar irradiation received by the planet (see legend).}
    \label{fig:metallicity_radius}
\end{figure}

\cref{fig:metallicity_radius} shows the radius vs.~planetary bulk metallicity for each mass at a fairly young (0.5 Gyrs) and old (5 Gyrs) age. Naively, one would expect the largest planet consists of pure H-He, and that the radius decreases monotonically as a function of bulk metallicity. This is indeed true when all the heavy elements are assumed to be in the center (dashed lines) and the envelope is composed only of H-He. It is also the case when the atmospheric temperature is low due to low stellar irradiation (dark blue lines), because for low temperatures the gas opacity does not scale strongly with the metallicity. However, when the heavy elements are mixed (solid lines) and there is significant stellar irradiation (dark and light green lines), the radius is no longer monotonically decreasing with $Z$!

The $Z = 0$ planet does not yield the largest radius, because there is a competition between the increase in density with larger $Z$ and the delayed cooling due to the higher opacity. At a given mass and age, below $Z \leq 0.1$ the effect of the opacity dominates, and the slower cooling is sufficient to increase the radius with metallicity. At $Z \simeq 0.1$, the radius peaks, and then decreases monotonically. Note that the turnover can occur at different bulk metallicities: in a) and b) at $Z \simeq 0.08$, in c) at $Z \simeq 0.05$ and in d) at $Z \simeq 0.06$. The exact location of the turnover depends on the details of the model (e.g., the competition between the internal flux and the stellar irradiation), and not simply on the planetary mass. One reason for this is because the metallicity-scaling of the opacity depends on the atmospheric temperature. We have already noted that in previous works core-envelope models yielded larger planets compared to fully-mixed models \citep{Baraffe2008,Vazan2013,Thorngren2016}. This is because these calculations did not account for the effect of the heavy elements on the opacity. The effect they found is small compared to the change in radius we observed caused by the metallicity-scaled opacity.

We therefore conclude that when the metallicity scaling of the opacity is considered, a pure H-He planet is not necessarily the largest. As a result, even if age, radius, mass and the heavy-element distribution of a planet were perfectly known, there would still be an ambiguity in the inferred heavy-element mass if the opacity or the heavy-element distribution are not well-determined. For example, it can be seen in \cref{fig:metallicity_radius} that for a 5 Gyrs old $1 M_J$ planet receiving a stellar flux of $10^{8}$ erg cm$^{-2}$ s$^{-1}$, either $Z = 0$ or $Z \simeq 0.15$ could fit the same radius. Clearly a detailed understanding of the atmospheric opacity is required.

\section{New Inflated Jupiters}\label{sec:inflated}
In this section, we compare our theoretical models with observed exoplanets from the T16 catalogue. We discuss the implications of the CMS H-He EoS for the mass-metallicity relation and show that the T16 sample includes inflated Jupiters.

\begin{figure}[h]
    \epsscale{1.1}
    \plotone{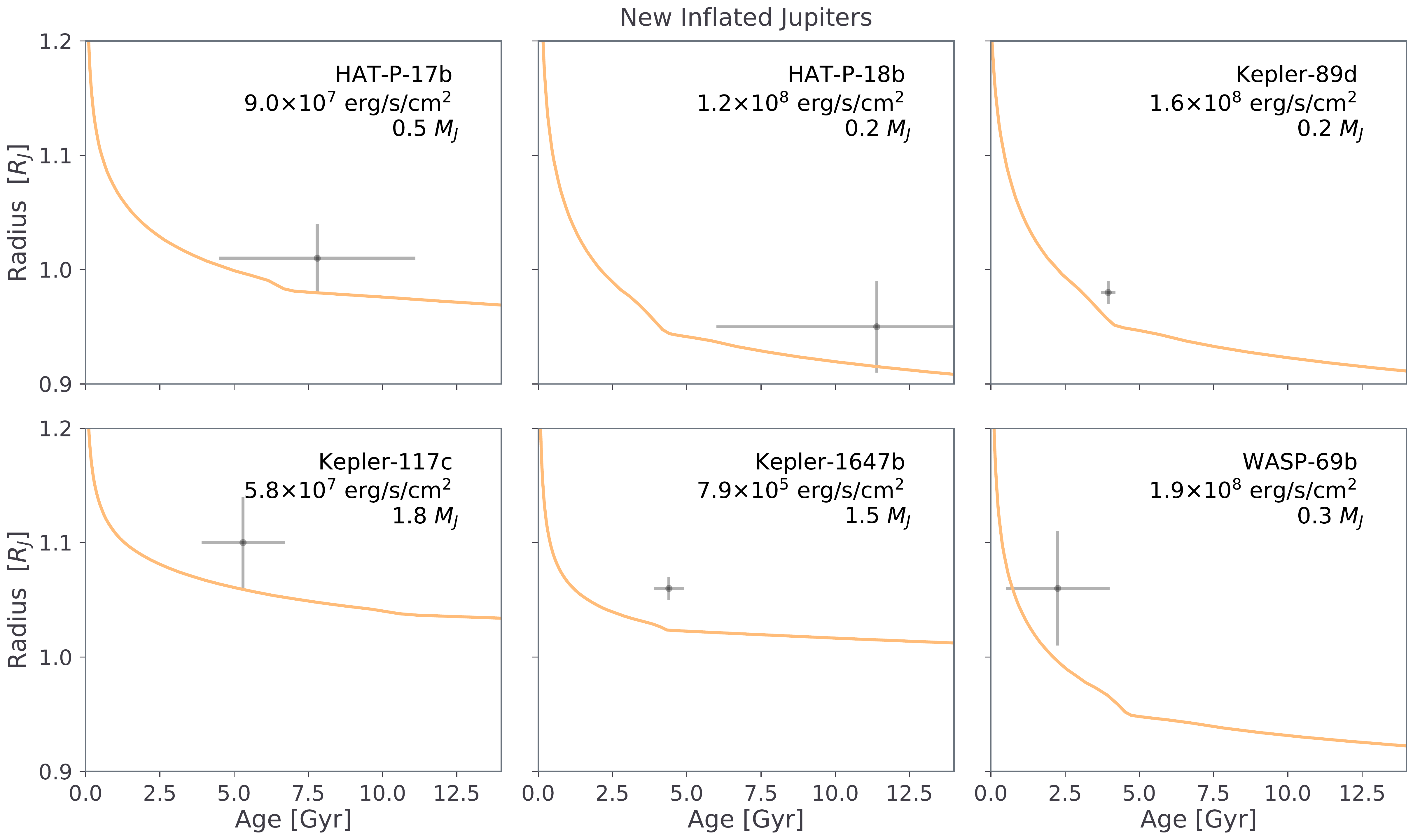}
    \caption{Radius evolution calculated for six different exoplanets assuming $Z = 0$ and a proto-solar hydrogen-helium ratio. The models account for the incident stellar irradiation. Black error bars show the measurement uncertainty of for a given planet. Each panel also lists the planet's name, insolation and mass.}
    \label{fig:too_small}
\end{figure}

As shown in \S \ref{sec:eos}, the radius of a planet for a given mass, bulk metallicity and age depends on the H-He EoS. We found that planets calculated with CMS instead of SCvH are smaller, with the effect being largest for low metallicity planets below a Jupiter mass. In order to investigate how the use of the CMS EoS affects the inferred mass-metallicity relation, we simulate the thermal evolution of the T16 planets, but this time using the CMS EoS. The other model parameters are the same as described in \S \ref{sec:methods} and \S \ref{sec:t16}: the planets are homogeneously mixed, and the opacity does not scale with the metallicity.

The results presented in \S \ref{sec:eos} indicate that using CMS leads to lower planetary metallicities.
In fact, for the T16 planets with inferred bulk metallicites of $Z \leq 0.1$ (their predictions) our calculations predict a median value of $Z = 0$ to fit these planets' radii. This was the case for 14 planets out of the 47 planets in the T16 catalogue (see \cref{tab:zeroed} in Appendix \ref{sec:appendix}). For 6 out of these 14 planets, we found that a pure H-He composition yields planets that are smaller than their observational measurements.

The radius evolution of these six planets is shown in \cref{fig:too_small}, together with the observational uncertainty on the age and radius. Note that while the observed radius is essentially described by a Gaussian distribution, the age is better described by a flat distribution, with each value being more or less equally likely (see \S 2 in \citet{Thorngren2016}).

We did not find any commonalities among these planets: their mass, stellar irradiation, stellar metallicity and stellar mass cover a large range with no discernible pattern (see \cref{tab:zeroed} in Appendix \ref{sec:appendix}). If one assumes that the CMS EoS yields a more realistic planetary model, then these planets are expected to be inflated giant planets.

Using the CMS EoS means roughly a quarter of the T16 planets are best fit with a bulk metallicity of zero. It is therefore challenging to derive a mass-metallicity relation, since the resulting slope depends on whether these planets are omitted or included when calculating the fit. If we exclude these planets, then our calculations yield roughly the same mass-metallicity trend of $M_z \propto M^{0.6}$ as inferred in T16. It is very unlikely that a giant planet has no metals, independent \edit1{of} whether it was formed by gravitational instability or core accretion \citep{Helled2014}.

Our results suggest that unless these planets are made of pure H-He, which is rather unlikely, they should be classified as inflated giant planets.

As discussed in \S \ref{sec:opacity_results} (see \cref{fig:metallicity_radius} in particular), one possible way to inflate these planets is by increasing their atmospheric opacity, for example due to a high atmospheric metallicity. Another way to enhance the opacity would be the inclusion of grains or clouds (see, e.g., \citet{Vazan2013,Poser2019} \edit1{and \S \ref{sec:observations} for a discussion)}.
\edit1{The planetary cooling can also be delayed if the planetary interior is not fully convective. Such planets could have a stably-stratified composition gradient in their deep interiors, where  heat transport is dominated by conduction or  layered convection. \citet{Kurokawa2015} investigated layered convection in the context of highly inflated hot Jupiters ($R \sim 2 R_J$). They find that while layered convection is insufficient  to explain the radii of these hot Jupiters, it could inflate planetary radii by $\sim$ 10\%.}
\edit1{It should be noted that opacity increase due to clouds or grains, or layered convection could be valid mechanisms to explain the inflated radii of planets shown in \cref{fig:too_small}}.
A more detailed investigation is beyond the scope of this paper, and we hope to address this in future research.

\section{Theoretical vs Observational Uncertainties}\label{sec:observations}
Our results show that the H-He EoS, the metallicity scaling of the opacity, and the distribution of the heavy elements have important effects on the inferred planetary radius and consequently on its bulk metallicity. Generally, the largest radii are inferred when using the SCvH EoS and assuming a fully mixed interior. The smallest sizes are obtained by using the CMS EoS with a core-envelope structure.

As observations are significantly improving it is important to have a better theoretical understanding of planet formation and evolution. It is therefore desirable to investigate how theoretical uncertainties compare to the observational ones, and whether they could be larger.

\begin{figure}[h]
    \epsscale{1.1}
    \plotone{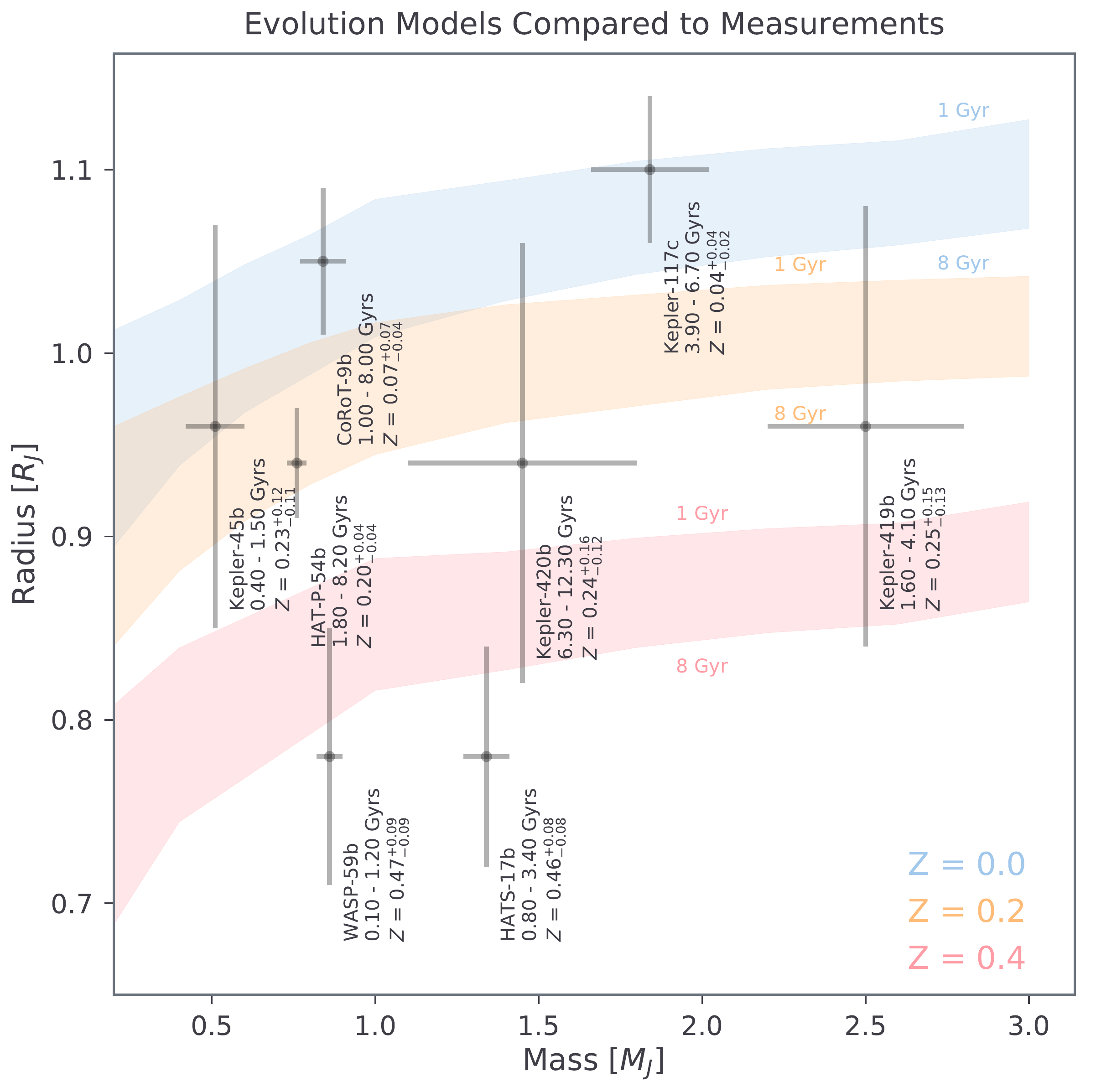}
    \caption{Mass-radius relation with observational uncertainties of 8 exoplanets. The coloured shaded regions are calculated with our thermal evolution models using three different bulk metallicities: $Z = 0$ (blue), $Z = 0.2$ (orange) and $Z = 0.4$ (pink), and indicate the the radius of the planet at a given mass from 1 Gyr (top boundaries) to 8 Gyr (bottom boundaries).}
    \label{fig:ages_zrange}
\end{figure}

In \cref{fig:ages_zrange}, we plot the mass-radius relation of a few selected exoplanets together with their observational uncertainties. The shaded regions were calculated with our evolution models for a specific bulk metallicity (Z = 0, 0.2 and 0.4). The extent of the shaded region corresponds to the calculated planetary radii between 1 to 8 Gyrs (top to bottom). This figure illustrates how large  measurement uncertainties in the age and radius translate into a large uncertainty in the inferred bulk metallicity. For planets with large associated observation uncertainties, the theoretical uncertainties induced by the model assumptions are not of great importance.

\begin{figure}
    \epsscale{0.8}
    \plotone{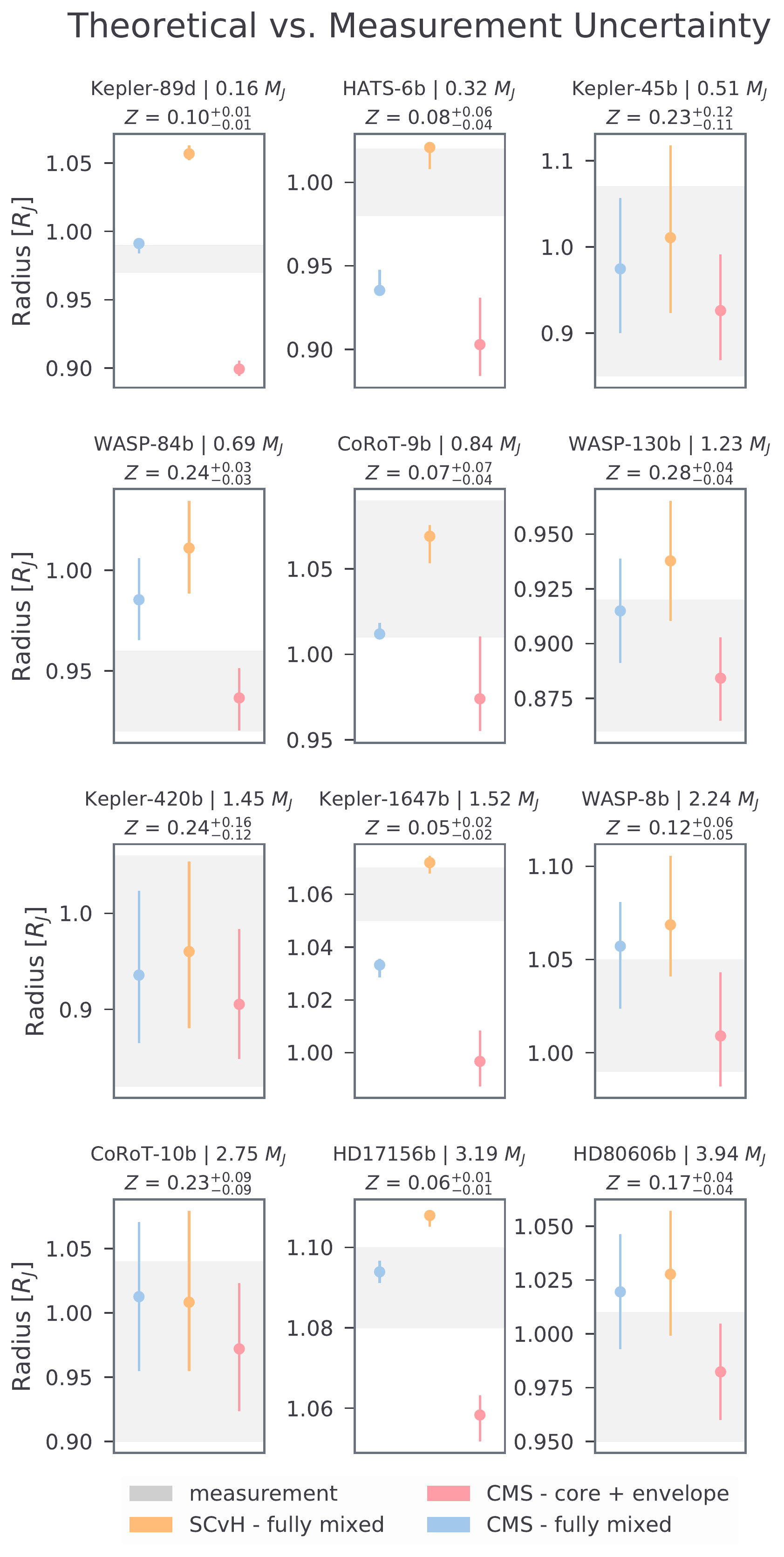}
    \caption{Exoplanetary radii from measurements and their uncertainties (shaded grey regions) together with our theoretical models (coloured lines and dots). The coloured dots are calculated from thermal evolution models for three different models: CMS EoS with a core-envelope structure (pink), CMS EoS homogeneously mixed (blue) and SCvH EoS homogeneously mixed (orange). The coloured error bars show the range in possible radii when using the range of bulk metallicities as inferred in T16. Radii are taken at the planet's median age, and the planet's mass and the median metallicity is given in the title of the panel. Note that our models use H$_2$O EoS for the heavy elements, and should not be compared directly to the observed radii, only to the size of the observational uncertainty.}
    \label{fig:err_cms_scvh}
\end{figure}

However, for well-constrained, low $Z$ planets like Kepler-89d or Hats-6b, it is interesting to compare how the differences in radii between the theoretical models compare to the measurement uncertainties. \cref{fig:err_cms_scvh} shows the radii of 12 of the 47 T16 exoplanets, together with the observational uncertainties (shaded grey rectangles). For each of these planets, we simulated the thermal evolution using the one-standard deviation range of bulk metallicities as inferred in T16. This was done for three model assumptions: (i) the CMS EoS with a fully mixed interior, (ii) the CMS EoS with a core-envelope structure, and the (iii) the SCvH EoS and a homogeneously mixed planet. The coloured error bars present the ranges of inferred radii. In these calculations we used the heavy-element EoS for water. This systematically over-estimates the radius in the models, since real planets are expected to also be composed of denser materials. A more realistic representation of the heavy elements would not change the differences between the models, since it would only result a shift which will be comparable in all cases. Therefore, the comparison of the uncertainty due to these model assumptions with measurement uncertainties is unaffected by the choice of the heavy-element EoS.

\edit1{Planets are unlikely to contain the same mixture of heavy elements. Therefore, the modeller's choice of the heavy-element EoS introduces another theoretical uncertainty, namely that all planets are calculated as if they have only one heavy element (or at most a mixture of two) and that it is the same in all planets. For example, the water-to-rock ratio is expected to be quite different depending on where in the disk the planet formed (due to different condensation temperatures). It also matters which material is used to represent rock, since their densities can be quite different. \citet{Baraffe2008} showed that changing the metal composition (from pure water to pure iron as an upper bound) produces differences on the order of 10\% in the radius for a  1 $M_J$ planet with $Z = 0.5$.}

\edit1{Additionally, we did not consider the effect clouds, grains or aerosols could have on the opacity. The effect of grains and grain vapour on the evolution of gas giants was investigated by \citet{Vazan2013}. For a 1 M$_J$ planet with a bulk metallicity of $Z = 0.2$, the additional opacity resulted in an increase of the planetary radius up to $\sim$ 10\% at a few Gyrs. Interestingly, this is a much larger effect than the one resulting from different heavy-element EoSs. \citet{Poser2019} investigated clouds as an additional opacity source, and found that optically thick clouds that are deep seated can also delay the cooling significantly. This effect also leads to $\sim$ 10\% increase in planetary radius.}
\edit1{These effects would only further increase the uncertainty in the planetary radius coming from theoretical models. In this work, we focused on several common assumptions and investigate them thoroughly. Therefore, the theoretical uncertainties we present here are lower estimates, and more realistic uncertainties are likely even larger.}

For a specific theoretical model the uncertainty in the planet's  radius translates to an uncertainty in the metallicity. However, the difference between the three theoretical models (colours) can be much larger. As mentioned above, this is significant for planets like Kepler-89d or HATS-6b that have small observational error bars. This clearly demonstrates that in many cases, the difference in radius between the different models is larger by a factor of a few than the observational uncertainties (see \cref{tab:uncertainties}). In other words, current theoretical uncertainties linked to the model assumptions overtake the observational ones. This means that in order to interpret future observations properly, progress in theory is required. This includes a better understanding of EoSs, opacities, as well as a linking of planet formation models to observations of the current-state properties.

\begin{table}[h]
    \begin{tabular}{@{}lcccc@{}}
    \toprule
        Planet & Mass [M$_J$] & Radius [R$_J$] & $\Delta R_{obs}$ [R$_J$] & $\Delta R_{th}$  [R$_J$] \\ \midrule
        Kepler-89d   & 0.16 & 0.98   & 0.02        & 0.17   \\
        HATS-6b      & 0.32 & 1.00   & 0.04        & 0.14   \\
        Kepler-45b   & 0.51 & 0.96    & 0.22        & 0.25   \\
        WASP-84b     & 0.69 & 0.94   & 0.04        & 0.11   \\
        CoRoT-9b     & 0.84 & 1.04   & 0.10        & 0.12   \\
        WASP-130b    & 1.23 & 0.89   & 0.06        & 0.10   \\
        Kepler-420b  & 1.45 & 0.94   & 0.24        & 0.21   \\
        Kepler-1647b & 1.52 & 1.06   & 0.02        & 0.09   \\
        WASP-8b      & 2.24 & 1.04   & 0.06        & 0.12   \\
        CoRoT-10b    & 2.75 & 1.04   & 0.14        & 0.16   \\
        HD17156b     & 3.19 & 1.09   & 0.02        & 0.06   \\
        HD80606b     & 3.94 & 0.98   & 0.06        & 0.10   \\ \bottomrule
    \end{tabular}
    \caption{Mass (error omitted), radius and the difference between the min. and max. radius $\Delta R_{obs}$ for the planets in \cref{fig:err_cms_scvh}. $\Delta R_{th}$ lists the uncertainty in planetary radius due to the different model assumptions as calculated by our thermal evolution models (see text for details).}
    \label{tab:uncertainties}
\end{table}

\section{Summary}\label{sec:summary}
In this paper we investigated the effect of \edit1{a few common} model assumptions such as EoS, opacity and distribution of heavy elements on the radius of giant exoplanets. We modelled the planetary thermal evolution, and showed that the various model assumptions can have large effects on the inferred radii of gaseous planets. 

We confirmed the mass-metallicity trend reported in \citet{Thorngren2016} when using the SCvH H-He EoS and QEoS for SiO$_2$, assuming the opacity does not scale with the heavy elements. However, when removing a 20 $M_{\oplus}$ heavy-element core from each planet, there is no longer a mass-metallicity correlation. Also, changing the H-He EoS from SCvH  \citep{Saumon1995} to CMS \citep{Chabrier2019} resulted in inferring a zero bulk metallicity for roughly 25\% of the planets from the \citet{Thorngren2016} sample. All of these planets had to be excluded when calculating the mass-metallicity trend, which resulted in a less statistically significant mass-metallicity relation. We conclude the choice of EoS, opacity, and distribution of heavy elements are important for the determination of the metallicity of gas giant planets.

\clearpage
Our main results can be summarized as follows:

\begin{enumerate}
    \item The CMS H-He EoS leads to smaller planets when compared to using SCvH. The effect can be as large as a 10\% change in radius, depending on the mass and bulk metallicity of the planet.
    \item A consequence of using the new H-He EoS is that six of the planets from the T16 catalogue are inflated beyond what is expected for metal-free gas giants (accounting for the irradiation and age uncertainty). This introduces new inflated warm gas giants. 
    \item Considering the gas opacity-enhancement due to heavy elements in the envelope of irradiated planets is important. The increased gas opacity at a given planetary mass and age leads to an increase of  $\simeq$ 10\% in radius. In some cases the radius does not monotonically decrease with bulk metallicity.
    \item Although the heavy-element distribution has a relatively small effect on the radius of gas giants, the change in gas opacity can change the radius significantly.
    \item The uncertainty caused by the different model assumptions considered here can be comparable or even larger than the observational uncertainties.
    \edit1{\item Clearly, there are other model assumptions that we do not consider in this work. This strengthens even further the importance of modeling for data interpretation.}
\end{enumerate}

Our study suggests that there is clear a connection between giant planet characterization and origin. Giant planet formation models can determine the expected primordial internal structures and therefore guide the model assumptions. Since the heavy-element distribution impacts the inferred radius (mostly via the opacity), it is valuable to understand when giant planets are expected to be fully mixed and under what conditions a core-envelope structure is more realistic.

It is clear that an improved understanding of giant exoplanets requires progress in both theory and observations. Accurate determinations of the stellar (and therefore planetary) age is expected from the Plato 2.0 mission. This will allow us to better constrain the planetary composition and internal structure of young giant exoplanets. In addition, the James Webb Space Telescope and Ariel missions will provide measurements of atmospheric compositions of gaseous planets. This, together with accurate measurements of the planetary mass, radius, and age can further constrain the planets' compositions and current-state structures, and therefore improved our understanding of giant planet formation and evolution.

Finally, progress in theory is also required. It is important to identify the H-He EoS that is the most appropriate for giant planet interiors, and to have more realistic representations of the heavy elements. It is also crucial to improve the opacity calculations and atmospheric models for highly irradiated planets and include non-solar chemical compositions as well as the effect of grains and clouds. The development of improved theoretical models will allow us to better interpret the upcoming exciting exoplanetary data. 

\acknowledgments{}

We thank the anonymous referee for thoroughly reading our paper and providing many useful comments and suggestions. We acknowledge support from SNSF grant \texttt{\detokenize{200020_188460}} and the National Centre for Competence in Research ‘PlanetS’ supported by SNSF.

% \newpage
\appendix
\section{Composition Gradients}\label{sec:comp_gradients}
\edit1{
In this work, we assumed that the heavy-elements are either homogeneously mixed in the envelope or are all in the core. Formation models suggest that giant planets could form with composition gradients, such that the heavy-element fraction decreases from the core towards the envelope \citep[e.g.,][]{Helled2017, Valletta2020}. 
However, it is unclear whether these composition gradients can be sustained over $10^9$ years \citep{Vazan2015,Mueller2020}, because convective mixing can erase such gradients. However, the outcome strongly depends on the initial condition, mixing parameters and other model assumptions. 
A full investigation of this topic is beyond the scope of this work. Nevertheless, in this section we show that the core-envelope and fully-mixed cases bracket the planetary radius through the evolution.}

\edit1{
We select Kepler-45b ($M = 0.51 \, M_J$, $Z = 0.25$) and Wasp-130b ($M = 1.23 \, M_J$, $Z = 0.30$) with these bulk metallicities. For each of the planets, we created three starting models with different initial radii $R_0$. The initial radii for Kepler-45b were: $R_0$ = 2.3 (Case 1), 2.0 (Case 2) 1.5 (Case 3), and for Wasp-130b: $R_0$ = 3.0 (Case 1), 2.4 (Case 2) and 1.9 (Case 3). These correspond to initial envelope entropies of $S_0$ = 9.6, 9.5, 9.0 erg g$^{-1}$ K$^{-1}$ (Kepler-45b) and $S_0$ = 10.9, 10.4, 9.8 erg g$^{-1}$ K$^{-1}$ (Wasp-130b). At the beginning of the simulation, we impose a composition gradient such that the heavy-element fraction linearly decreases from the center towards the surface. This means that there is an entropy gradient for roughly 60 \% of the planet by mass, and the entropies quoted above are only valid for the outer, homogeneously mixed part of the envelope. The planets are then evolved until the upper-bound estimate of their age (1.4 Gyrs for Kepler-45b and 14 Gyrs for Wasp-130b). Based on the Ledoux criterion, the energy transport in the interior can be either by radiation/conduction or by convection. Material is allowed to be mixed in convective regions, and convection is treated as a diffusive process with the mixing length theory  \cite[e.g.,][]{Kippenhahn2012}.}

\begin{figure}
    \epsscale{1.15}
    \plotone{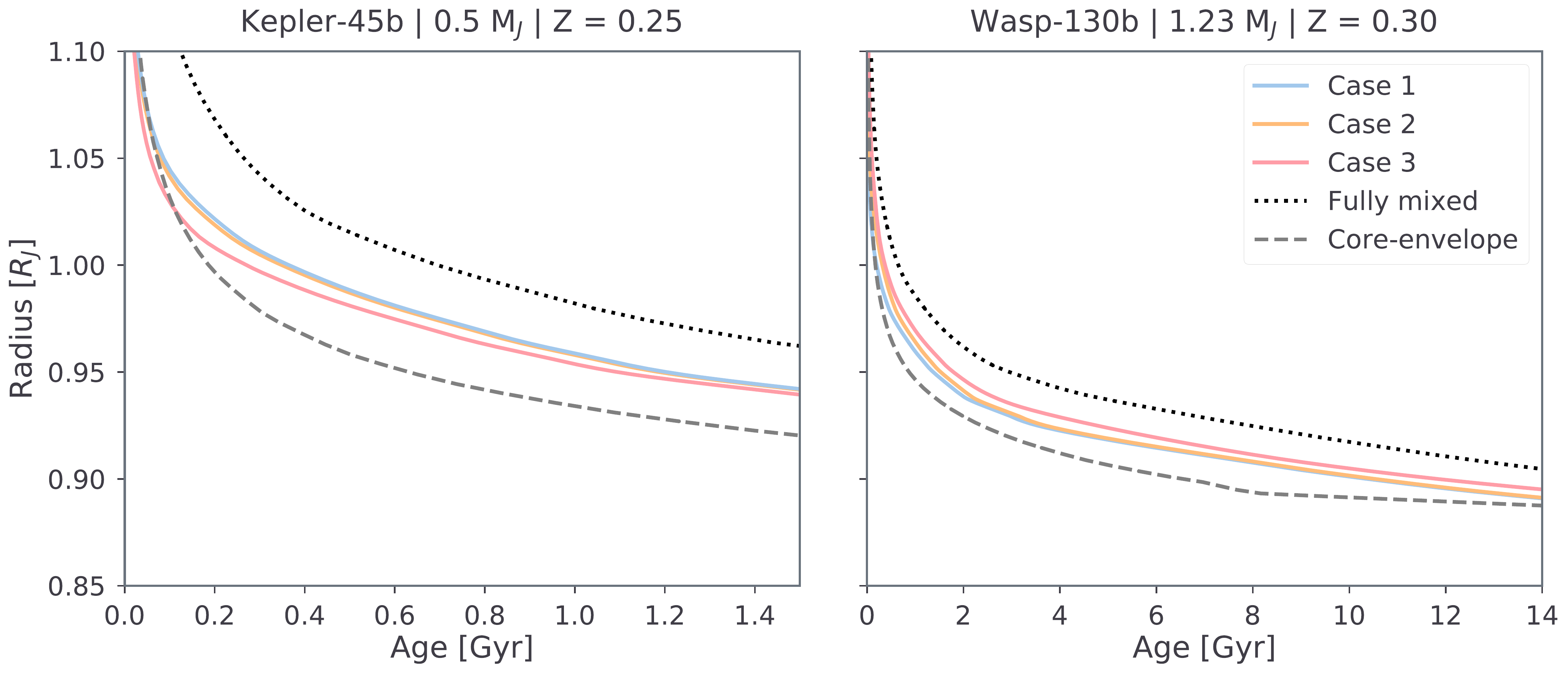}
    \caption{Radius evolution of Kepler-45b with $Z = 0.25$ (left) and Wasp-130b with $Z = 0.30$ (right). The coloured lines show the models with the primordial composition gradient for three different initial conditions (Case 1, 2 and 3 - see text for details). For comparison, the fully-mixed (dotted black) and core-envelope (dashed gray) structures are also shown.}
    \label{fig:kep45b_wasp130b_radius}
\end{figure}

\edit1{
The radius evolution of both planets is plotted in \cref{fig:kep45b_wasp130b_radius} for the different initial conditions (coloured lines). 
The radii of the planets with the composition gradients are between the radii of the fully-mixed and core-envelope cases. This demonstrates that they provide reasonable upper and lower bounds for the radius at a given age. There are plausible scenarios where this may not be the case: (i) if layered-convection operates, the composition gradient could remain stable for Gyrs and delay the cooling significantly and (ii) if the planet is very young ($\lesssim 10^{7}$ yrs), large scale convection may still be inhibited by the stabilizing gradient \citep[e.g.,][]{Mueller2020}.}

\section{Additional table}\label{sec:appendix}

\begin{table}[H]
    \centering
    \begin{tabular}{@{}lcccccc@{}}
    \toprule
        Planet & Mass [M$_J$] & Radius [R$_J$] & Age [Gyr] & Flux [erg cm$^{-2}$ s$^{-1}$] & [Fe/H] \\ \midrule
        Corot-9b     & $0.84 \pm 0.07$ & $1.05 \pm 0.04$ & 1.0 - 8.0 & $6.59 \times 10^{6}$ & $-0.01 \pm 0.06$ \\
        HAT-P-15b    & $1.95 \pm 0.07$ & $1.07 \pm 0.04$ & 5.2 - 9.3 & $1.51 \times 10^{8}$ & $0.22 \pm 0.09$ \\
        HAT-P-17b    & $0.53 \pm 0.02$ & $1.01 \pm 0.03$ & 4.5 - 11.1 & $8.97 \times 10^{7}$ & $0.00 \pm 0.08$ \\
        HAT-P-18b    & $0.18 \pm 0.03$ & $0.95 \pm 0.04$ & 6.0 - 16.8 & $1.18 \times 10^{8}$ & $0.10 \pm 0.08$ \\
        HATS-6b      & $0.32 \pm 0.07$ & $1.00 \pm 0.02$ & 0.1 - 13.7 & $5.84 \times 10^{7}$ & $0.20 \pm 0.09$ \\
        HD17156b     & $3.19 \pm 0.03$ & $1.09 \pm 0.01$ & 2.4 - 3.8 & $1.98 \times 10^{8}$ & $0.24 \pm 0.05$ \\
        Kepler-30c   & $2.01 \pm 0.16$ & $1.10 \pm 0.04$ & 0.2 - 3.8 & $1.12 \times 10^{7}$ & $0.18 \pm 0.27$ \\
        Kepler-75b   & $10.10 \pm 0.40$ & $1.05 \pm 0.03$ & 3.4 - 9.7 & $1.29 \times 10^{8}$ & $0.30 \pm 0.12$ \\
        Kepler-89d   & $0.16 \pm 0.02$ & $0.98 \pm 0.01$ & 3.7 - 4.2 & $1.57 \times 10^{8}$ & $0.01 \pm 0.04$ \\
        Kepler-117c  & $1.84 \pm 0.18$ & $1.10 \pm 0.04$ & 3.9 - 6.7 & $5.79 \times 10^{7}$ & $-0.04 \pm 0.10$ \\
        Kepler-432b  & $5.84 \pm 0.05$ & $1.10 \pm 0.03$ & 2.6 - 4.2 & $1.73 \times 10^{8}$ & $-0.02 \pm 0.06$ \\
        Kepler-1647b & $1.52 \pm 0.65$ & $1.06 \pm 0.01$ & 3.9 - 4.9 & $7.91 \times 10^{5}$ & $-0.14 \pm 0.05$ \\
        WASP-69b     & $0.26 \pm 0.02$ & $1.06 \pm 0.05$ & 0.5 - 4.0 & $1.94 \times 10^{8}$ & $0.14 \pm 0.08$ \\
        WASP-80b     & $0.54 \pm 0.04$ & $1.00 \pm 0.03$ & 0.5 - 10 & $1.06 \times 10^{8}$ & $-0.13^{+0.15}_{-0.17}$ \\ \bottomrule
    \end{tabular}
    \caption{Exoplanets for which our models predict a median planetary bulk metallicity of $Z = 0$ when using the CMS H-He EoS. Listed are their names, mass, radius, age, received stellar irradiation and stellar metallicity. See \S \ref{sec:inflated} for details.}
    \label{tab:zeroed}
\end{table}

\newpage
\bibliography{library}
\bibliographystyle{aasjournal}

\end{document}